\documentstyle[psfig,aps,prl,amsfonts,amssymb]{revtex}

\begin{document}
\title{Analyses of a Yang-Mills Field 
over the Three-Level Quantum Systems}
\author{Paul B. Slater}
\address{ISBER, University of
California, Santa Barbara, CA 93106-2150\\
e-mail: slater@itp.ucsb.edu,
FAX: (805) 893-7995}

\date{\today}

\draft
\maketitle
\vskip -0.1cm

\begin{abstract}
Utilizing a number of results of Dittmann, we investigate the nature of
the Yang-Mills field over the eight-dimensional convex set, endowed with the
Bures metric, of three-level quantum systems.
Adopting a numerical strategy, we first decompose the field into 
self-dual and anti-self-dual components, by implementing the octonionic
equations of Corrigan, Devchand, 
Fairlie and Nuyts. For each of these 
three fields, we obtain approximations to: (1) the Yang-Mills
functional; (2) certain 
quantities studied by Bilge, Dereli, and
Ko\c cak in their analysis of self-dual Yang-Mills fields in eight
dimensions; and (3) other measures of interest.
\end{abstract}

\vspace{.2cm}
\hspace{1.5cm} Keywords: Bures metric, Yang-Mills fields, 
three-level quantum systems, action,

\vspace{.005cm}

\hspace{3.3cm} numerical integration 
\hspace{1.5cm}

\vspace{.15cm}

\hspace{1.5cm} Mathematics Subject  Classification (2000): 81T13, 58Z05


\vspace{.1cm}

The Bures metric, defined on the nondegenerate density matrices, has been the
object of considerable study 
\cite{uhlmann,hubner1,hubner2,brauncaves,slaterexact}. 
It is the {\it minimal} member of
the nondenumerable family of {\it monotone} metrics \cite{petzsudar,rusles}.
Other members of this family of particular note are the 
``Bogoliubov-Kubo-Mori'' (BKM) 
metric \cite{michor,grasselli}, the maximal monotone metric, as well as
the ``Morozova-Chentsov''  \cite{slatclarke} and 
``quasi-Bures'' \cite{slaterhall} ones, 
the last yielding the minimax/maximin asymptotic
redundancy in universal quantum coding \cite{kratt}.
Interestingly, these (operator) monotone 
metrics correspond in a direct fashion
 to certain
``measures of central tendency'', with, for 
example,  the Bures metric corresponding
to the {\it arithmetic} mean, $(x+y)/2$,  of numbers $x$ and $y$
\cite{petzsudar,rusles}.
All these monotone metrics constitute various extensions to the quantum
domain of the (unique) Fisher information metric in the classical realm
where the objects of study are probability distributions
(rather than density matrices)
\cite{kass} (cf. \cite{frieden}). However,
the only one of these monotone metrics  
 that can be extended to the boundary of pure states
yielding the standard Fubini-Study metric on this boundary
 is the Bures (minimal
monotone) one \cite[sec. IV]{petzsudar}.

Dittmann has shown that ``the connection form (gauge field) 
related to the generalization of the Berry phase to the mixed states
proposed by Uhlmann satisfies the source-free Yang-Mills equation 
$*D*D \omega$, where
the Hodge operator is taken with respect to the Bures metric on the space
of finite-dimensional density matrices'' \cite{ditt1} (cf. \cite{rudy}).
(Let us also indicate here that the BKM metric has been shown to be the
unique monotone Riemannian metric for which two ``natural flat''
connections --- the ``exponential'' and ``mixture'' ones --- are
dual \cite{grasselli}.)
Here we report our efforts to evaluate --- using numerical 
(lattice-gauge-like \cite{gockeler}) methods --- the
action functional for this  Yang-Mills field over the 
eight-dimensional convex set ($M$) 
equipped with the Bures metric of $3 \times 3$ density matrices, as well
as for the two self-dual components of this field. (Using an exact version for 
the $2 \times 2$ density matrices of the methodology described below, 
we have found a value of precisely three
 for the action of the corresponding Yang-Mills field.
The Chern-Simons functional is zero.)

To proceed, we exploit our recent work 
\cite{slaterjgp1,slaterjgp2} in determining the elements of the
Bures metric using a certain Euler angle parameterization of the
$3 \times 3$ density matrices \cite{byrdslater}. 
(Earlier, Dittmann had noted \cite{ditt1} ``that in affine coordinates
(e. g. using the Pauli matrices for $n = 2$) the [Bures] metric 
becomes very complicated for $n>2$ and no good parameterization seems to
be available for general $n$''.) 
This allows us, among other
things, to employ as our 
parameter space four (noncontiguous)  eight-dimensional hyperrectangles
(each having three sides of length $\pi$, three sides of length 
${\pi \over 2}$,  one  ${\pi \over 4}$ in length and one 
$\cos^{-1}{{1 \over \sqrt{3}}}$), rather than less 
analytically convenient ones, such as that discussed by
Bloore \cite[Fig. 3]{bloore}, involving spheroids and ``tetrapaks''.
(The four hyperrectangles are not adjacent due to the fact that two
of the six Euler angles used in the parameterization have disconnected ranges.
This  situation --- which results in different normalization factors
 only --- has only come to our attention relatively recently, and 
serves as an erratum to earlier analyses \cite{markbyrd1,markbyrd2}, 
which had relied upon work of Marinov \cite{marinov1}, without taking into
account a correction  \cite{marinov2} to \cite{marinov1} (cf. \cite{caves}).)

To be specific, we seek to implement the formula \cite[(1.1.10)]{tian}
for the Yang-Mills functional $YM(A)$ of the 
appropriate connection $A$
\begin{equation} \label{YM1}
YM(A) = {1 \over 4 \pi^2} \int_{M} |F_{A}|^{2}\mbox{d} V_{g}.
\end{equation}
Here $F_{A}$ is the curvature of $A$ 
and, for our purposes, $\mbox{d} V_{g}$ is the volume form of
the Bures metric $g$ on the eight-dimensional manifold $M$, composed of the
three-level quantum systems.
More detailedly \cite{tian},
\begin{equation}
|F_{A}|^{2} = \sum_{i,j,k,l} <F_{\alpha i j},F_{\alpha k  l} >   g^{i k} 
g^{j l},
\end{equation}
where $(g_{ij})$ is the metric tensor of $g$ and $(g^{ij})$ is its 
inverse. (Formulas for $V_{g}$, ($g_{ij}$) and ($g^{ij}$), using the
Euler-angle parameterization \cite{byrdslater}, have been derived
in \cite{slaterjgp1} (cf. \cite{slaterhall}. Let us indicate here that since
we directly adopt formulas from a number of different sources, our overall
notation may not be as uniform as fully desirable, though hopefully 
readily enough comprehensible to the informed reader.)

To achieve the decomposition $F_{A} = F_{A}^{+} + F_{A}^{-}$
of $F_{A}$ into the sum of  ``anti-self-dual'' ($F_{A}^{+}$) and 
``self-dual'' ($F_{A}^{-}$) parts \cite[eq. (2.9)]{gao} \cite[eq. (3.4)]{bs}, 
we use the formulas first given
by Corrigan {\it et al} \cite{corrigan}, and presented again numerous times
({\it e. g.} \cite{baulieu,singer,bilge}).
The components of the anti-self-dual field satisfy a set of seven 
linear equations 
(corresponding to an eigenvalue 1) \cite[eq. (3.39)]{corrigan} and those of 
the self-dual
field, a set of twenty-one linear equations 
(corresponding to an eigenvalue of -3) \cite[eq. (3.40)]{corrigan}.
The set  of seven equations has been viewed as more fundamental in nature
than the set of twenty-one equations by Figueroa-O'Farrill,
who is also somewhat critical of the application
 of the terms ``self-dual'' and
``anti-self-dual'' in this octonionic setting \cite{FF}.

For the required connection form $A$, Dittmann has presented the 
general formula
\cite[eq. (9)]{ditt1}
\begin{equation} \label{tsd}
A = {1 \over \tilde{L} +\tilde{R}} (W^{*} T - T^{*} W)
\end{equation}
Here the elements of $T$ lie  in the tangent space to the 
principal $U(\mathcal{H})$-bundle, the manifold of invertible normalized 
($\mbox{Tr} W^{*} W = 1$) Hilbert-Schmidt operators, while 
$L$ is the operator (depending on $W$) of left multiplication by
the density matrix 
$\rho = W W^{*}$ and $R$ the corresponding operator of right multiplication.
Also, $\tilde{L}$ and $\tilde{R}$ are the counterpart operators
for $\tilde{\rho} =W^{*} W $.
Since the Euler-angle parameterization of the $3 \times 3$ density 
matrices presented in \cite{byrdslater} is of the (``Schur-Schatten'')
form, 
\begin{equation}
\rho = U D U^{*}, 
\end{equation}
$U$ being  unitary, $U^{*}$ its conjugate 
transpose, and $D$ 
the diagonal matrix composed 
of the three eigenvalues of $\rho$, one can immediately express $W$ as
$U D^{1/2} U^{*}$, with $W=W^{*}$.

In the non-trivial task of implementing formula (\ref{tsd}) 
for the three-level
quantum systems, we relied upon the implicit relation 
(equivalence) between two 
formulas for the Bures metric 
for $n$-level quantum systems \cite[eqs. (2) and (16)]{ditt2},
\begin{equation}
g={1 \over 2} \mbox{Tr} \mbox{d} \rho {1 \over L + R} \mbox{d} \rho,
\end{equation}
and
\begin{equation}
g = {1 \over 2} \sum^{n}_{i,j} a_{ij} \mbox{Tr} \mbox{d} \rho \rho^{i-1} 
\mbox{d} \rho \rho^{j-1}.
\end{equation} (The somewhat involved formula for the coefficients
$a_{ij}$, functions of elementary invariants, is stated in Proposition 3
of \cite{ditt2}.)
In other words, we used the formula for the connection
\begin{equation}
A = {1 \over 2} \sum^{3}_{i,j} a_{ij} \rho^{i-1} S \rho^{j-1},
\end{equation}
where $S = W^{*} T - T^{*} W$.
(We were able to express some of the 
individual components of $A$ in relatively simple
forms, but others remain, at this stage, quite cumbersome in nature.)

Let us note that in  a previous 
similarly-motivated study \cite{slaterjgp2} to that here, 
we pursued many of the same
questions, but  
somewhat misguidedly employed for our
analyses the curvature ($R$) of the Bures 
Riemannian metric, rather than the curvature ($F_{A}$)
of the gauge field $A$, so the interpretation to be given to the results 
there, in retrospect, is
 not altogether clear. (By embedding the spin connection in the gauge 
connection, one can construct a self-dual gauge field directly from
a self-dual metric \cite{acharya}.) In that analysis \cite{slaterjgp2}, 
numerical evidence led us to conclude that for the $8 \times 8$
skew-symmetric matrices of the $R_{abcd}$, holding $c,d$ constant, 
one of the four pairs of imaginary eigenvalues is 
always degenerate, that is (0,0) cf. \cite{gilkey1,gilkey2}.

In \cite{slaterjgp2}, we pursued both a Monte Carlo numerical integration
scheme, randomly selecting points lying in 
the ranges of the eight parameters, as well as a scheme in 
which the points were systematically selected as the grid points 
of a hyperrectangular lattice. Here we 
have chosen to focus on a lattice scheme, employing
$1,024 = 4 \times 256$
grid points (cf. \cite{gockeler}).
 The coordinates of the nodes of the lattice
were chosen in the natural (symmetric) manner, so as to divide each 
individual side of the four  hyperrectangles into three intervals, the middle
interval being twice the length of the two end ones, which under wrapping
(identifying the end points)
form  a single interval equal in length to the middle one.
Based upon this lattice scheme, we approximated 
the value of the Yang-Mills functional (\ref{YM1}) for the Yang-Mills 
field $F_{A}$ itself to 
be 137.653, for the ``anti-self-dual'' part
$F_{A}^{+}$ to be 222.806, and for the ``self-dual'' part $F_{A}^{-}$ to be 
176.194 (Table I).

\begin{table}
\begin{tabular}{r||c|c|c|}
quantity & $F_{A}$ & $F^{+}_{A}$ & $F^{-}_{A}$ \\
\hline \hline
Yang-Mills functional & 137.653 & 222.806 & 176.194 \\
\hline
$\int_{M} (F,F)^{2}$ & 37.2769 & 15.0131 & 1.1757 \\
\hline
$\int_{M}(F^2,F^2)$ & 162.468  & 97.6351 & 4.77715 \\
\hline
$\int_{M} (\mbox{tr} F^2,\mbox{tr} F^2)$ & 129.285 & 72.9629 & 8.54855 \\
\hline
$H_{4} $ & $ -1.35 \cdot 10^{-11}$ & 4.15623 &  35.6896 \\
\hline
$H_{2}  $ & $ -1.74 \cdot 10^{-11}$   & 5.4683 &  71.7151 \\
\hline
$H_{4}   - H_{2}/2$ & 
$-4.77 \cdot 10^{-12}$ & 1.42208  & -.167908 \\
\end{tabular}
\label{pop}
\caption{Lattice approximations, 
based on $1,024 = 4 \times 2^{8}$ grid points, 
to various quantities of interest
for the eight-dimensional Yang-Mills field $F_{A}$ over the three-level 
quantum systems, and for its
anti-self-dual ($F_{A}^{+}$) and self-dual ($F_{A}^{-}$) constituents}
\end{table}

The explanation for the 
Yang-Mills functionals being {\it larger}  for both $F_{A}^{+}$ and 
$F_{A}^{-}$ than for $F_{A}$ itself, in apparent contradiction 
to the general assertion in \cite[sec.2]{gao},
may be that the topological types of $F_{A}^{+}$ and
$F_{A}^{-}$ are 
different from that of $F_{A}$. (I thank Y.-H. Gao for suggesting this 
possibility.)  It should be noted, though, that 
in more than four dimensions, 
the Yang-Mills functional itself need not be a minimum 
of the action  
\cite[sec. 2.4]{bais} \cite{ivanova}.

Motivated by  the work of Bilge, Dereli and Ko\c cak (BDK) \cite{bilge} 
concerning $SO(n)$-bundles in eight-dimensional spaces, 
we computed the integrals over $M$
of $(F_{A},F_{A})^2$, $(F_{A}^{2},F_{A}^{2})$ and  
$(\mbox{tr} F_{A}^{2},\mbox{tr} F_{A}^{2}$), and also similarly for the
self-dual and anti-self-dual parts of $F_{A}$.
BDK considered a generic action density of the form
\begin{equation}
a (F_{A},F_{A})^2 + b (F_{A}^2,F_{A}^2) +
 c (\mbox{tr} F_{A}^2, \mbox{tr} F_{A}^2).
\end{equation}
(The action density proposed by Grossman, Kephardt and Stasheff 
\cite{grossman} corresponds to the choice $a=c=0$.)
BDK derived a topological bound
\begin{equation}
\int_{M} (F_{A},F_{A})^2 \geq k \int_{M} p_{1}(E)^{2}
\end{equation}
that is achieved by ``strongly (anti)self-dual'' Yang-Mills fields
(those conforming to the set of twenty-one --- not the set of 
seven --- equations of
Corrigan {\it et al}), where 
$p_{1}$ is the first Pontrjagin class of the $SO(n)$ Yang-Mills
bundle $E$ and $k$ is a constant.
This is to be compared with a topological bound obtained earlier
\cite{grossman}
\begin{equation}
\int_{M} (F_{A}^2,F_{A}^2) \geq k' \int_{M} p_{2}(E).
\end{equation}

We also approximated for the three fields ($F_{A}, F^{+}_{A}$ and $F^{-}_{A}$)
the quantities
\cite[eq. (A1)]{tchrakian}, where
(using  $\tilde{F}$ to generically represent $F_{A}$, $F_{A}^{-}$ and
$F_{A}^{+}$)
\begin{equation}
H_{4}  = \int_{M} 
\epsilon_{\mu \nu \rho \sigma \tau \lambda 
\kappa \eta} \mbox{tr} (\tilde{F}_{\mu \nu} 
\tilde{F}_{\rho \sigma}\tilde{F}_{\tau \lambda} 
\tilde{F}_{\kappa \eta})
\end{equation}
and
\begin{equation}
H_{2}  =  \int_{M} 
\epsilon_{\mu \nu \rho \sigma \tau \lambda \kappa \eta} \mbox{tr} 
(\tilde{F}_{\mu \nu} \tilde{F}_{\rho \sigma})
\cdot \mbox{tr} (\tilde{F}_{\tau \lambda} \tilde{F}_{\kappa \eta}),
\end{equation}
for which $H_{4} - H_{2}/2$ is proportional to a certain topological
invariant, $q_{8}$ \cite[eq. (A1)]{tchrakian}
($\epsilon_{\mu \nu \rho \sigma \tau \lambda \kappa \eta}$ 
is the completely antisymmetric Levi-Civita symbol).
Let us note that Table I indicates that for 
the Yang-Mills field $F_{A}$, the terms $H_{4}$
and $H_{2}$ are both essentially zero. ($H_{i}$ is proportional to the
$i$-th Chern number. ``The Chern numbers are those obtained by integrating
characteristic polynomials of degree dim $M$ over the entire manifold $M$''
\cite[Def. (2.5.2)]{acz}.)

In contrast to the Yang-Mills functional (\ref{YM1}), we see 
from Table I that 
(following the work of BDK \cite{bilge}) the integrals over the Riemannian
manifold composed of the 
three-level quantum systems endowed with the Bures metric
of $(F,F)^2$, $(F^2,F^2)$ and $(\mbox{tr} F^2, \mbox{tr} F^2)$ are largest
for the Yang-Mills field $F_{A}$  \cite{ditt1}, smaller for the
anti-self-dual part $F_{A}^{+}$ and smallest for the self-dual component
$F_{A}^{-}$.

In the context of $SO(n)$-bundles and 2-forms in 
$2 m$ dimensions represented by $2 m \times 2 m$ skew-symmetric 
real matrices, BDK derived the result that
\begin{equation}
(F_{A},F_{A})^2 \geq {2 \over 3} (\mbox{tr} F_{A}^2, \mbox{tr} F_{A}^2).
\end{equation}
We see from Table I that for our $SU(3)$-bundle, on the other 
hand, this result does not appear to hold for all three fields.

For analyses of non-self-dual Yang-Mills fields (in four dimensions)
see \cite{sibner,wang,sadun,bor} and (more generally, in $2 n$ dimensions)
\cite{burzlaff} (cf. \cite{duff}).

The computational demands in deriving (with the use of MATHEMATICA) the
results reported above in Table I have been considerable
 --- due, in particular, to the still somewhat cumbersome formulas for 
certain of 
the elements of the Bures metric and the term $W^{*} T - T^{*} W$ 
[occurring in 
the formula (\ref{tsd}) for the connection $A$]  that 
we have derived \cite{slaterjgp1,slaterjgp2}, using the parameterization
in \cite{byrdslater}.
To employ instead of the $1,024 = 4 \times 2^{8}$
point 
lattice, a $26,244 =4 \times 3^{8}$ one, for example, presently seems too
much for us to reasonably accomplish. (Actually, it appears that since the
results for the four noncontiguous hyperrectangles are all identical,
we would only have to undertake $6,561 = 3^{8}$ independent evaluations.)
However, we have been able to refine the lattice in {\it two} of the eight
directions, so that we deal with $2,304 = 4 \times 3^2  \times 2^6$
points. If we choose these two directions
(more or less arbitrarily)  to correspond to one of the six
Euler angles ($\alpha$ in the notation of \cite{slaterjgp1,slaterjgp2})
and one of the spherical angles parameterizing the eigenvalues
($\theta_{1}$),
then we obtain the outcomes reported in Table II.
\begin{table}
\begin{tabular}{r||c|c|c}
quantity & $F_{A}$ & $ F^{+}_{A}$ & $F^{-}_{A}$ \\
\hline \hline
Yang-Mills functional & 197.458  & 324.53 &  256.239 \\
\hline
$\int_{M} (F,F)^{2}$ & 37.9854  & 15.3356  & 1.20481 \\
\hline
$\int_{M} (F^{2},F^{2})$ & 162.107  & 98.5983  & 4.92389 \\
\hline
$\int_{M} (\mbox{tr} F^2,\mbox{tr} F^2)$ & 132.557 & 74.8023 & 8.77179 \\
\hline
$H_{4}$ & -$3. 08 \cdot 10^{-12}$ & 4.20952 & 36.6592 \\
\hline
$H_{2}$ & $4.13 \cdot 10^{-11}$ & 5.28634 & 73.5915 \\
\hline
$H_{4} -H_{2}/2$ & $-2.37 \cdot 10^{-11}$  & 1.56635 & -.136522  \\
\end{tabular}
\label{pip}
\caption{Lattice approximations, based on 
$2,304 = 4 \times 3^{2} \times 2^{6}$ grid points, to various quantities
of interest for the eight-dimensional Yang-Mills field $F_{A}$ 
over the three-level quantum systems, and for its
anti-self-dual ($F_{A}^{+}$) and self-dual ($F_{A}^{-}$)
constitutents}
\end{table}
We see (quantitative, not really qualitative) changes in Table II 
from the values reported in Table I, witt quite substantial increases in
the case of the Yang-Mills functionals, and comparatively 
modest differences in the other sets of indices \cite{bilge}.

\acknowledgments

I would like to express appreciation to the Institute for Theoretical
Physics for computational support in this research.

\end{document}